\documentclass[useAMS]{mn2e}
\usepackage{psfig}

\usepackage{times}
\def\spose#1{\hbox to 0pt{#1\hss}}
\newcommand\lsim{\mathrel{\spose{\lower 3pt\hbox{$\mathchar"218$}}
     \raise 2.0pt\hbox{$\mathchar"13C$}}}
\newcommand\gsim{\mathrel{\spose{\lower 3pt\hbox{$\mathchar"218$}}
     \raise 2.0pt\hbox{$\mathchar"13E$}}}
\def\ltsima{$\; \buildrel < \over \sim \;$}
\def\lsim{\lower.5ex\hbox{\ltsima}}
\def\gtsima{$\; \buildrel > \over \sim \;$}
\def\gsim{\lower.5ex\hbox{\gtsima}}

\def\sch{Schwarzschild}

\title[Blue  FSRQs]
{Blue {\it Fermi} Flat Spectrum Radio Quasars}

\author[Ghisellini et al.] 
{G. Ghisellini$^1$\thanks{E--mail:gabriele.ghisellini@brera.inaf.it}, F. Tavecchio$^1$, 
L. Foschini$^1$, T. Sbarrato$^{2,1}$, G. Ghirlanda$^1$, L. Maraschi$^3$ 
\\
1: INAF -- Osservatorio Astronomico di Brera, via E. Bianchi 46, I--23807 Merate, Italy \\
2: Universit\'a dell'Insubria, via Valleggio 11, 22100 Como, Italy\\
3: INAF -- Osservatorio Astronomico di Brera, via Brera 28, I--20100 Milano, Italy \\
}

\begin{document}


\pagerange{\pageref{firstpage}--\pageref{lastpage}} \pubyear{2007}

\maketitle

\label{firstpage}

\begin{abstract}

Many blazars detected by the {\it Fermi} satellite, observed spectroscopically
in the optical, are line--less, and have been classified as BL Lac objects.
Optical--UV photometry of nearly one hundred of them 
allowed to determine the redshift for a handful of objects
and redshift upper limits for the great majority. 
A few of these are candidates to be ``blue quasars", namely flat spectrum radio quasars
whose  broad emission lines are hidden by an overwhelming synchrotron emission peaking in the UV.
This implies that the emitting electrons have high energies.  
In turn, this requires relatively weak radiative cooling, a condition that can be 
met if the main radiative dissipation of the jet power occurs outside the broad line region.
We confirm this hypothesis by studying and modelling the spectral energy  
distributions of the 4 ``blue quasars" recently discovered.
Furthermore, we discuss the distribution of {\it Fermi} blazars
in the $\gamma$--ray spectral index -- $\gamma$--ray luminosity plane,
and argue that ``blue quasars" objects are a minority within the blazar populations.
\end{abstract}


\section{Introduction}

The  {\it Fermi} satellite is detecting $\gamma$--ray emission from 
a large number of blazars (Flat Spectrum Radio Quasars, FSRQs, and BL Lacs).
From the data collected in the first two years of operation a  ``clean"  sample was constructed 
from the data of the {\it Fermi}/Large Area Telescope (LAT) instrument, the 2LAC sample 
(Ackermann et al. 2011), that allows population studies. 

The 2LAC sample includes 395 sources classified as ``BL Lacs", 310 FSRQs, 157 sources of 
``unknown type", 4 Narrow Line Seyfert 1 (Abdo et al. 2009)
and other 18 ``non blazar" AGNs and 2 starburst galaxies.
Of the 395 sources classified as BL Lacs, 56\% lack a  redshift determination, which limits the 
possibility of discussing  their physical properties.
When an emission lines is visible, the subdivision between the BL Lac and FSRQs categories is based
on the equivalent width (EW) of the line, as measured in the rest frame: the blazar
is classified as BL Lac if the rest frame EW of any permitted line is smaller than 5 \AA\ 
(Stickel et al. 1991).

In order to gain redshift information for BL Lacs 
without any visible line, 
Rau et al. (2012, hereafter R12)
set up a program to obtain simultaneous photometry over a wide wavelength range using the Gamma--Ray Burst 
Optical/Near--Infrared Detector (GROND) and the {\it Swift}/Optical Ultraviolet Telescope (UVOT).

80 blazars with optical/radio identification but without
redshift information were selected from the 2LAC sample based on celestial position and 
small foreground reddening; 8 more (2 with known redshift) were included  though not part of 
the clean 2LAC sample because of Fermi data quality problems; 
other 16 2LAC sources with known redshift were included for verification.
In total, 104 blazars were considered: 82 have been classified as 
BL Lacs by Ackermann et al. (2011), 3 as FSRQs and 19 as of unknown type.

All the sources (but one, due to a lack of precise coordinates)
were observed simultaneously with GROND and with the {\it Swift} UVOT. 
GROND can observe simultaneously
in 7 filters, from 2.4 $\mu$m ($K$ band) to
4000 \AA\ ($g$ band), while UVOT can observe in 6 filters 
(one filter at the time), 
from 5400 \AA\ ($v$ band) to $\sim$ 1600 \AA\ ($uvw2$ filter; with center
wavelength at 2000 \AA).
These data, covering the wavelength interval from  2.4 $\mu$m to 1600 \AA,  allow
to derive photometric redshifts if the source is far enough to be affected by
intervening  Ly$\alpha$ absorption.

As a result, photometric  redshifts were determined at the 90\% confidence level for
11 sources, including 3 blazars that also had a previously measured spectroscopic redshift
(in one case it is only a lower limit on $z$, derived by the presence 
of an intervening absorption line).
Furthermore, the absence of any Ly$\alpha$ absorption feature in the spectrum
could lead to the estimate of an {\it upper limit} on the redshift for 81 blazars,
including 12 blazars with known spectroscopic redshift.
The upper limits in these 12 sources were all consistent with the spectroscopic measurement.

Among the 11 blazars with $z$ estimated photometrically, 
Padovani, Giommi \& Rau (2012, P12 hereafter) discussed 4 sources, chosen
because of the flat (in $\nu F_\nu$) optical continuum 
(i.e. $F_\nu \propto \nu^{-\alpha}$ with $\alpha \le 1$).
They constructed their spectral energy distributions (SEDs), including
the X--ray data of the X--Ray Telescope (XRT) onboard the {\it Swift} satellite
and the $\gamma$--ray data of {\it Fermi}.
They concluded that,  despite their featureless optical spectrum and  high synchrotron  peak frequency, 
more typical of low power BL Lac objects (if blazars obey the ``blazar sequence" 
as proposed by Fossati et al. 1998), these sources are probably FSRQs 
whose broad emission lines are swamped by the beamed non--thermal continuum.

The aim of the present  paper is twofold.
The first aim is to analyze in detail the SEDs of  the 4 blazars discussed in P12
in order examine whether the large $\gamma$--ray luminosity together with  a high frequency
synchrotron peak is in contrast with the physical interpretation
of the blazar sequence given in Ghisellini et al. (1998).

This interpretation 
assumes that  the peak frequency of the blazar synchrotron emission
depends mainly on the energy of the emitting electrons which, in turn,
depends on the rate of radiative cooling. 
In this scheme, we expect that high power blazars (i.e. FSRQs) 
emit most of their radiation within the broad line region (BLR), which 
can be the dominant source of seed photons for scattering. 
Radiative cooling, dominated by the Inverse Compton process on this external
population of seed photons 
(i.e. seed photons produced {\it externally} to the jet,
so the inverse Compon process is called {\it external Compton,}
EC for short) is severe, and therefore the energy of the
relevant electrons, emitting at the peak of the spectral 
energy distribution (SED), is small.
At the other extreme of the blazar sequence we have low power BL Lacs,
that lack (or have weak) broad emission lines.
EC is much less important, and the corresponding cooling is unimportant.
Electrons can then reach high energies, and produce high synchrotron 
(and self Compton, SSC) frequencies.

However, as described in Ghisellini \& Tavecchio (2008, see also Georganopoulos et al. 2001), 
intermediate situations may exist
if the dissipative region of the jet lies {\it beyond} the BLR.
In this case the relativistic electrons are not subject to strong EC losses,
and the main emission processes become synchrotron and SSC, 
with some contribution from EC scattering of IR photons produced by the torus.
The resulting SED is then foreseen to be similar to the SED of classical BL Lac objects.
It is then possible to have a {\it ``blue" quasar}  i.e. an object  
with emission lines and at the same time a SSC--dominated jet.
These conditions should occur in sources  with a relatively small accretion disc 
luminosity $L_{\rm d}$ and/or a relatively large black hole mass $M$.

The second aim of this paper is to take advantage of the upper limits in redshift,
made possible by the combination of GROND and UVOT observations,
to locate these blazars in the $\alpha_\gamma$--$L_\gamma$ plane,
to see if they violate the general trend observed in 
Ghisellini, Maraschi \& Tavecchio (2009).
In that paper, we analyzed only the blazars with spectroscopic redshift
detected in the first 3 months of {\it Fermi}.
We can now update the sample using the 2LAC 
catalog and including
blazars with redshift information provided by R12.

We use a flat cosmology with $H_0 =70$ km s$^{-1}$ Mpc$^{-1}$,  $\Omega_{\rm M}$=0.3
and the notation $Q=10^X Q_X$ in cgs units.

\section{Analysis of {\it Swift} data}

P12 presented the data of the 4 blazars in their sample only in the form of 
SED plots.
Since the {\it Swift} data are publicly available on the {\it Swift} archive,
we have analyzed both the XRT and UVOT data of the in order to check the results of P12.

 The data were screened, cleaned and analysed with the software package
HEASOFT v. 6.12, with the calibration database updated to 22 March 2012.
The XRT data were processed with the standard procedure ({\texttt{XRTPIPELINE v.0.12.6}). 
All sources were observed in photon counting (PC) mode and grade 0--12 
(single to quadruple pixel) were selected. 
The channels with energies below 0.3 keV and above 10 keV were excluded from the fit 
and the spectra were rebinned in energy so to have at least 20--30 counts per bin
in order to apply the $\chi^2$ test. 
When there are no sufficient counts,
we applied the likelihood statistic as reported by Cash (1979). 
Each spectrum was analysed through XSPEC v. 12.7.1
with an absorbed power law model with a fixed Galactic column density as
measured by Kalberla et al. (2005). 
The computed errors represent the 90\% confidence interval on the spectral parameters. 
The X--ray spectra displayed in the SED have been properly rebinned to ensure the best 
visualization.

UVOT (Roming et al. 2005) source counts were extracted from 
a circular region 5"--sized centred on the source position, 
while the background was extracted from 
an annulus with internal radius of 7" and variable outer radius 
depending on the nearest contaminating source.
Data were integrated with the \texttt{uvotimsum} task and then 
analysed by using the  \texttt{uvotsource} task.
The observed magnitudes have been dereddened according to the formulae 
by Cardelli et al. (1989) and converted into fluxes by using standard 
formulae and zero points.

The {\it Swift} data analysed by us are the same used in R12 and P12.
We found substantial agreement in three cases, but for the blazar
RX J0035.2+1515 we found UVOT fluxes rather different from R12,
and for this source we found no break.
Note that in the field of RX J0035.2+1515 there is a very bright star at a distance 
of 36" (TYC 1187--1355--1, with $B$=10.39)  
that causes problems when subtracting the background.
For its estimation, we have used an annular region of 
size 7"--20" centered on the blazar
(see also \S 4.1 below).

\section{BL Lacs or Flat Spectrum Radio Quasars?}

The 4 considered blazars have a featureless optical continuum,
and can be classified as BL Lacs if the classical definition is adopted
(i.e. an equivalent width of the lines less than 5 \AA).
However, P12 suggested that these blazars are instead FSRQs, whose emission
lines are swamped by the relativistically boosted jet flux.
The main argument for this classification is the strong radio power of these sources,
that is typical of FSQRs.
We agree with this interpretation, and we would like to offer 
another argument in favor of the FSRQ classification of these blazars.

Ghisellini et al. (2011) and Sbarrato et al. (2012), 
considering {\it Fermi} detected blazars, have shown that
there is a correlation between the $\gamma$--ray luminosity and
the luminosity of the broad lines, that includes not only FSRQs,
but also sources that are classified as BL Lacs according to the
classical subdivision (based on the EW of the lines).
Furthermore, if the luminosities are measured in Eddington units,
there is a divide between BL Lacs and FSRQs for 
$L_{\rm BLR}/L_{\rm Edd}\sim 5\times 10^{-4}$ and for
$L_\gamma/L_{\rm Edd}\sim 0.1$.
The latter values is obtained using the isotropic equivalent
of the $\gamma$--ray luminosity, i.e. the K--corrected $\gamma$--ray flux
multiplied by $(4\pi d^2_{\rm L})$, where $d_{\rm L}$ is
the luminosity distance. 
This of course does not imply that $L_\gamma$ is isotropic.
If the photometric redshift of our 4 blazars is correct, then their
$\gamma$--ray luminosity is large, exceeding the $L_\gamma/L_{\rm Edd}\sim 0.1$
value even for black hole masses equal to $M=10^9 M_\odot$.
An exception could be RX J0035.2+1515, if the true redshift is $z\sim 0.3$,
since in this case $L_\gamma\sim 10^{45}$ erg s$^{-1}$, making
$L_\gamma/L_{\rm Edd}< 0.1$ for $M> 8\times 10^7 M_\odot$.
In this case we should consider the source as a BL Lac, namely
a blazar whose emission lines, if present, are intrinsically weak.

The correlation found in Sbarrato et al. (2012) concerns
sources closely aligned with the line of sight, and it is
foreseen that when the $\gamma$--ray sensitivity will improve,
what we see now as a correlation is in fact a boundary
in the $L_{\rm BLR}$--$L_\gamma$ plane.
Bearing this in mind, the correlation has the form
\begin{equation}
L_{\rm BLR} \sim 4 \, L_\gamma^{0.93}
\label{tullia}
\end{equation}
with a large scatter, since the $\gamma$--ray luminosity is
highly variable in single objects even when averaging over one or two years
(see e.g. Ghirlanda et al. 2011).
This offers a rough way to estimate the luminosity of the broad lines.
When a good optical spectrum is available, we can then suggest
the minimum ratio -- between the boosted non--thermal and the thermal continua --
needed to hide the lines.

\section{Notes on individual sources}

\subsection{RX J0035.2+1515}  

The source has been observed
by the SDSS to have a featureless continuum.
NED reports $z=1.09$. 
On the other hand SDSS reports $z=1.057$ 
as a result of an automatic analysis, and also alerts that $z$ is actually unknown.
The quoted values are not believable.
Adopting the photometric redshift given in R12, $z=1.28$,
we can derive a (5$\sigma$) upper limit on the flux of the MgII line 
which is the most prominent broad line observable in the
spectral range of SDSS.
We  derive $L_{\rm MgII} < 1.4\times 10^{43}$ erg s$^{-1}$.
We then use the template given in Francis et al. (1991),  
adding the H$\alpha$ contribution (not included in Francis et al. 1991), 
with a relative weight of 77 (on a scale in which the  Ly$\alpha$ is 100).
The total weight of all lines is then 555 (see Celotti, Padovani \& Ghisellini et al. 1997),
and the weight of MgII is 34.
Therefore we derive $L_{\rm BLR} = (555/34) L_{\rm MgII} < 3\times 10^{44}$
erg s$^{-1}$.
With a covering factor of 0.1, the upper limit on the accretion disc
luminosity is $L_{\rm d}< 3\times 10^{45}$ erg s$^{-1}$.
The 0.1 value for the covering factor is uncertain and
should be taken as an average value with some dispersion
(see e.g. Baldwin \& Netzer 1978; Smith 1981).

Assuming a standard, geometrically thin optically thick disc (Shakura \& Sunyaev 1973), the 
peak frequency of its spectrum has $\nu_{\rm p} L_{\rm d}(\nu_{\rm p}) < 1.5 \times 10^{\rm 45}$
erg s$^{-1}$,  a factor $\sim 20$ below
the observed $\nu L_\nu$ in the optical, that has a luminosity 
$\sim 3\times 10^{46}$ erg s$^{-1}$.
We do not have any telling observation for the value of the black hole mass.
Sbarrato et al. (2012), studying a sample of FSRQs detected by {\it Fermi},
calculated the average black hole mass of FSRQs: $\langle M\rangle =5\times 10^8 M_\odot$.
Each black hole mass was estimated by Shen et al. (2011) through virial methods.
Using this average black hole mass, the ratio $L_{\rm d}/L_{\rm Edd} < 0.05$.

As long as $L_{\rm d}\gsim 10^{-2}\, L_{\rm Edd}$, the 
hypothesis of a standard disc is justified (radiatively inefficient disc should 
corresponds to Eddington ratios smaller than 0.01). 

As discussed in \S 3, the correlation between the BLR luminosity and the 
$\gamma$--ray luminosity in the {\it Fermi}/LAT
energy band offers a way to estimate $L_{\rm BLR}$ and $L_{\rm d}$.
Using Eq. \ref{tullia} and 
$L_\gamma \sim 4\times 10^{46}$ erg s$^{-1}$, we obtain 
$L_{\rm BLR} \sim 10^{44}$ erg s$^{-1}$ (and $L_{\rm d} \sim 10^{45}$ erg s$^{-1}$)
with an uncertainty of at least a factor 4.
Reassuringly, this estimate is consistent with the value found above.

The radio information are poor, since only the 1.4 GHz NRAO VLA Sky Survey (NVSS)
point is available, with a flux of 18.7 mJy (Condon et al. 1998)
and $\nu L_\nu \sim 2\times 10^{42}$ erg s$^{-1}$.
The source is detected in the infrared by the WISE satellite in all
its four filters (3.4, 4.6, 12 and 22 $\mu$m)\footnote{Cutri et al. 2012:
http://wise2.ipac.caltech.edu/docs/release/allsky/}.
The corresponding data points are consistent with the extrapolation of the 
spectrum derived from the GROND fluxes.

We have re--analyzed the UVOT data, finding a very bright source
at $\sim 36"$ from RX J0035.2+1515, as mentioned in \S 2.
Estimating the background in a region of the sky free of sources, we
have derived de--reddened fluxes 
significantly smaller than the ones reported
in R12, and a harder spectrum, with no sign of a break.
There is then the possibility that the derived photometric redshift
is affected by the uncertainties caused by incorrectly subtracting
the background. 
For this reason, we will consider for this source both the 
photometric redshift derived by R11, and also $z=0.3$.
This roughly corresponds to the lower limit on $z$ due to the
non--detection of the host galaxy
 both in the image and in its possible contribution to the SED
(see Wagner et al. 1996; Sbarufatti, Falomo \& Treves 2005).

{\subsection{SUMMS J053748--571828} 

Not observed by SDSS, its photometric redshift is  $z=1.55$.
To estimate the presence or not of a standard accretion disc,
and therefore its BL Lac or FSRQ nature, we can use the correlation
between $L_\gamma$ and $L_{\rm BLR}$, giving $L_{\rm BLR}\sim 6.3\times 10^{43}$ erg s$^{-1}$
and therefore an accretion disc 10 times more powerful.
Please note that the dispersion around the $L_\gamma$--$L_{\rm BLR}$ is large,
so the above values should be taken as an order of magnitude estimate.
Nevertheless, since the optical continuum in this source has a luminosity
similar to RX J0035.2+1515, it is conceivable that the synchrotron flux
has swamped the (indeed present) broad emission lines, if the optical spectrum has
a S/N similar to RX J0035.2+1515 (i.e. $\sim 40$), or worse.
The line of reasoning is the following:
assuming that the photometric redshift is correct,
we know at what frequencies the prominent broad emission line (i.e. Mg II) should appear.
In order to be visible, this line should have a minimum luminosity,
depending on the quality of the spectrum.
Since R12 state that the optical spectrum is featureless, 
we can then assign a lower limit on the line luminosity
assuming a S/N ratio. We find that a BLR of luminosity $\sim 6\times 10^{43}$ erg 
is consistent with the absence of lines in the optical spectrum if the S/N $<$40.

The source is detected by WISE. Although not simultaneous with GROND, 
the IR data points lie on the extrapolation of the spectrum defined
by the GROND points.

In the radio, the source flux is 99.8 mJy at 843 MHz, as reported
by the Sydney University Molonglo Sky Survey (SUMSS; Mauch et al. 2003).
This corresponds to $\nu L_\nu\sim 10^{43}$ erg s$^{-1}$,
slightly larger than RX J0035.2+1515 at 1.4 GHz.

\subsection{CRATES J0630--2406} 

Not observed by SDSS, its photometric redshift is  $z=1.6$.
The $L_\gamma$--$L_{\rm BLR}$ relation gives $L_{\rm BLR}\sim 5.5\times 10^{44}$ erg s$^{-1}$,
suggesting a rather luminous disc ($L_{\rm d}\sim 5.5\times 10^{45}$ erg s$^{-1}$, 
within on order of magnitude).
If true, this disc luminosity would correspond to a $\nu_{\rm p}L_{\rm d}(\nu_{\rm p})$ 
a factor 20 below the optical flux: a spectrum of the same quality of
the SDSS spectrum of RX J0035.2+1515 would not reveal any line.
The source is detected by WISE. The corresponding IR data well match the 
extrapolation of the spectrum defined by the GROND points.

The large radio luminosity, exceeding $10^{44}$ erg s$^{-1}$, 
indicates a powerful source, suggesting that this blazar is a FSRQ.

\subsection{CRATES 1312--2156} 

This source can be identified with PKS 1309--216.
The absorption line observed by Blades, Murdoch \& Hunstead (1980), identified as CIV,
gives a lower limit on the source redshift: $z>1.491$.
The radio luminosity is large, exceeding $10^{44}$ erg s$^{-1}$.
This suggests that this source is a FSRQ, whose accretion disc continuum
and lines are swamped by a strong synchrotron component.
We can estimate (approximately) an upper limit to the contribution of the 
disc, by assuming that, if observed with the SDSS, broad lines do not appear.
With a quality of the spectrum similar to the SDSS spectrum of RX J0035.2+1515,
and assuming the photometric redshift, the most prominent observable broad line 
would be MgII. 
Since it is not observed, the accretion disc must lie well below the observed
continuum, with a peak $\nu_{\rm p} L_{\rm d}(\nu_{\rm p})$ at least a factor $\sim 20$ below
the optical continuum, i.e. below $5\times 10^{45}$ erg s$^{-1}$.
This value is of the same order of what derived by the $L_\gamma$--$L_{\rm BLR}$ correlation,
that gives $L_{\rm d}\sim 6\times 10^{45}$ erg s$^{-1}$.
Also this source is detected by WISE. Although not simultaneous with GROND, 
the WISE data lie on the extrapolation of the spectrum defined
by the GROND points.

\begin{figure}
\hskip -0.3cm
\psfig{file=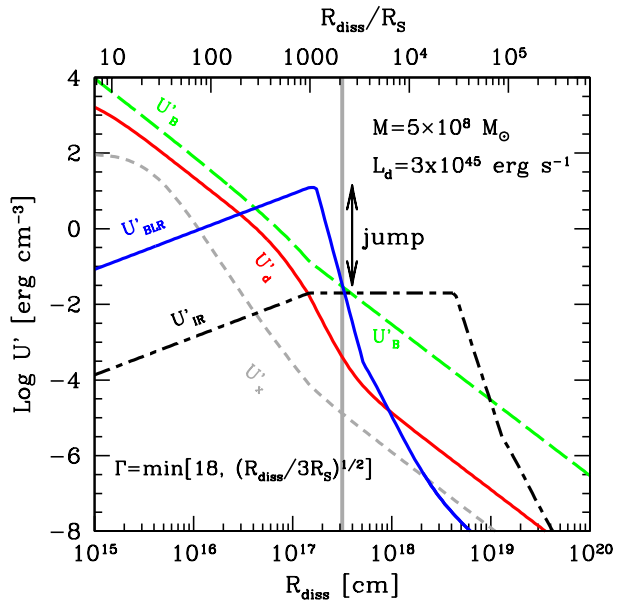,height=9cm,width=9cm}
\vskip -0.5 cm
\caption{ 
Energy densities as seen in the comoving frame
of a function of distance from the black hole, $R_{\rm diss}$.
The adopted mass is $M=5\times 10^8 M_\odot$ and the
disc luminosity is $L_{\rm d}=3\times 10^{45}$ erg s$^{-1}$.
These are the values used to model the SED of RX J0035+1515
when assuming $z$=1.28.
The vertical grey line marks the value of $R_{\rm diss}$
used in Fig. \ref{0035}.
The jet is assumed to accelerate with $\Gamma \propto (R_{\rm diss}/R_{\rm S})^{1/2}$
until it reaches its maximum value (in this case $\Gamma=18$).
At $R_{\rm diss} \gsim 1.7\times 10^{17}$ cm, the source is outside the BLR,
and at $R_{\rm diss}\sim 3\times 10^{17}$ cm
$U^\prime_{\rm BLR}$ is of the same order of $U^\prime_B$ and $U^\prime_{\rm IR}$,
if the IR radiation produced by a molecular torus intercepts 10\% of $L_{\rm d}$.
Further out, but within $R_{\rm IR}$, the radiation energy produced by the 
torus dominates again the radiative cooling, and the spectrum
is dominated again by the EC process.
The label "jump" indicates the jump in external radiation energy density 
as we move outside the BLR.
} 
\label{u0035}
\end{figure}

\begin{figure}
\vskip -0.5cm
\hskip -0.3 cm
\psfig{file=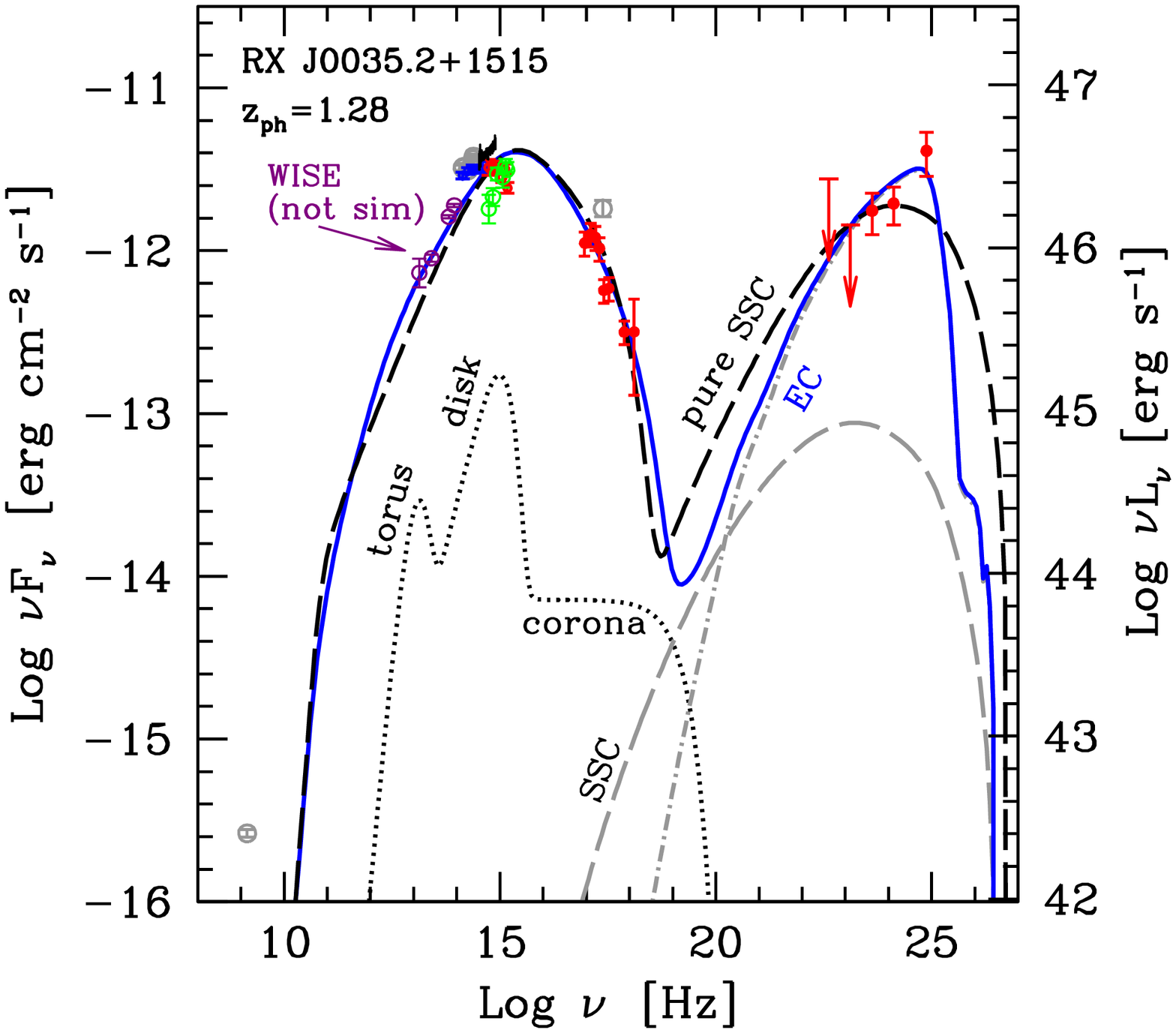,height=8cm,width=8.5cm}
\vskip -1 cm 
\hskip -0.3 cm
\psfig{file=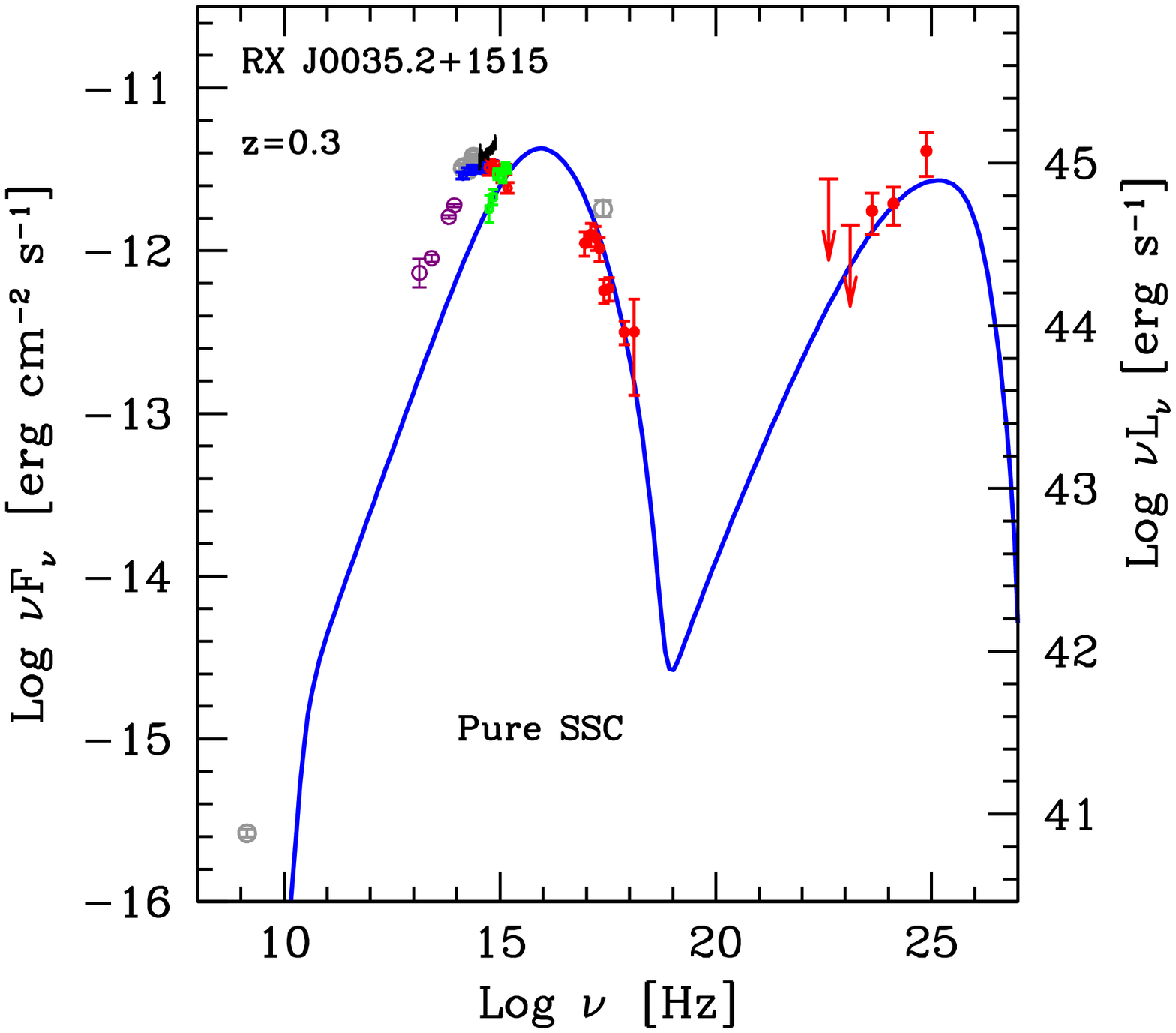,height=8cm,width=8.5cm}
\vskip -0.5 cm 
\caption{
Top panel:
The entire SED of 0035+1515, assuming $z_{\rm ph}$=1.28.
The (green) empty circles are the UVOT data as analyzed by us.
The black line in the optical is the (de--reddened) SDSS spectrum.
The WISE not simultaneous data (labelled ``WISE not sim") 
are also shown.
The solid line is a synchrotron plus EC model, the thick black dashed
line is a pure SSC model. 
The grey dotted line is the contribution of the assumed accretion disc, 
IR torus and X--ray corona, as labeled.
The dashed line is the contribution of the SSC flux for the
SSC+EC model.
Bottom panel: The SED assuming $z=0.3$.
The shown model is a pure SSC model,
assuming no external radiation.
}
\label{0035}
\end{figure}

\begin{figure}
\vskip -0.5cm
\hskip -0.3cm
\psfig{file=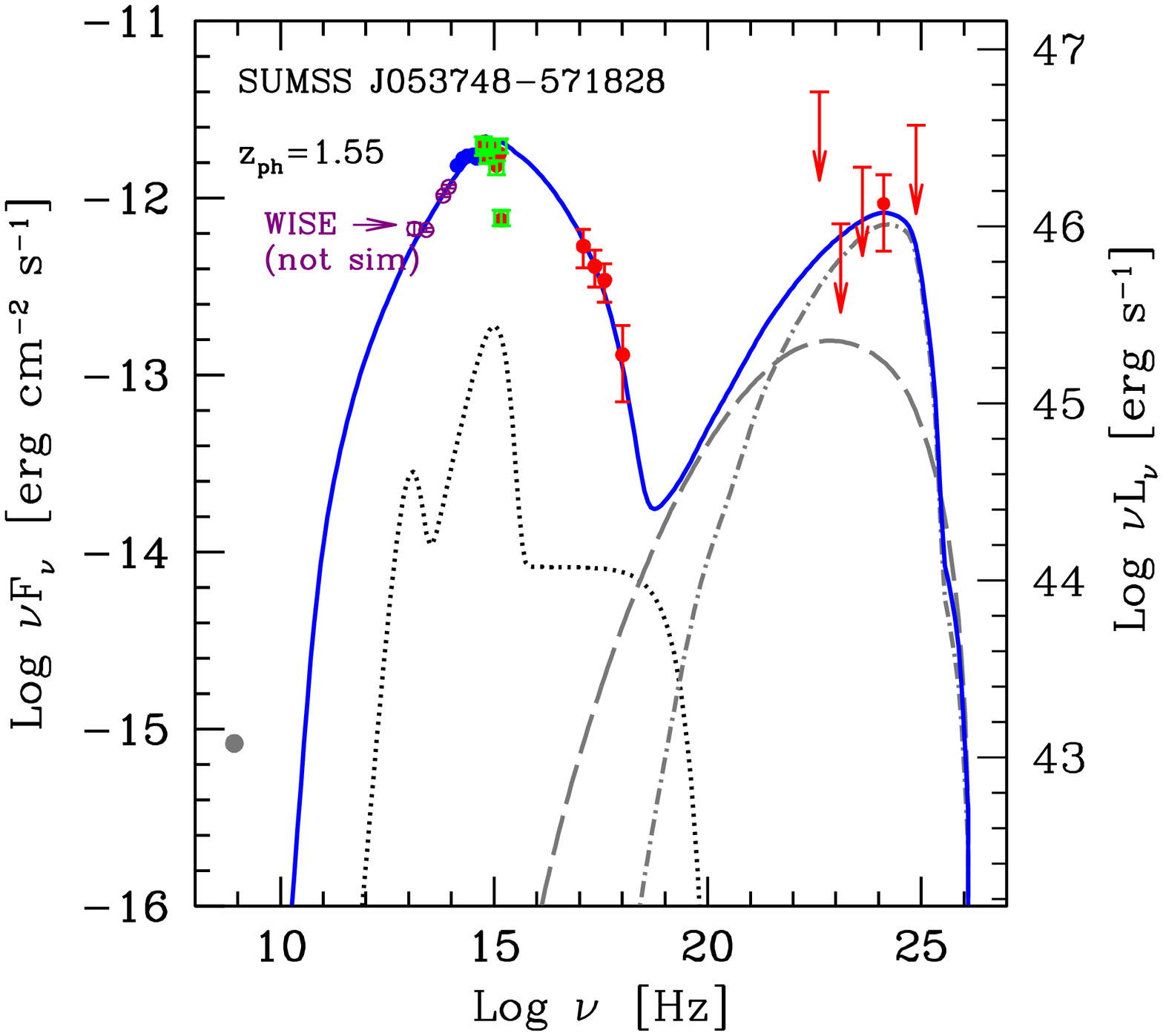,height=8cm,width=8.5cm}
\vskip -0.8cm
\hskip -0.3cm
\psfig{file=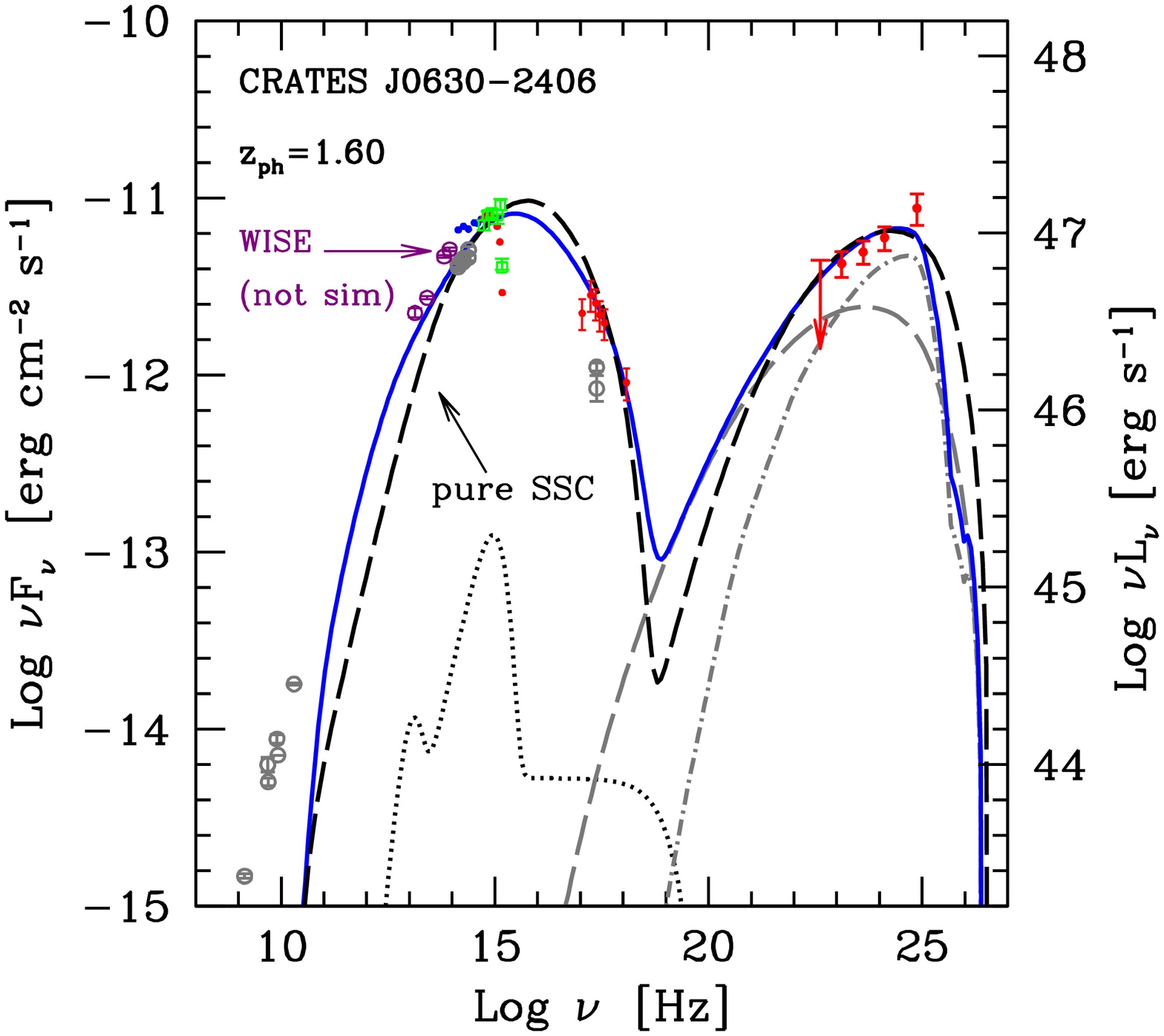,height=8cm,width=8.5cm}
\vskip -0.8cm
\hskip -0.3cm
\psfig{file=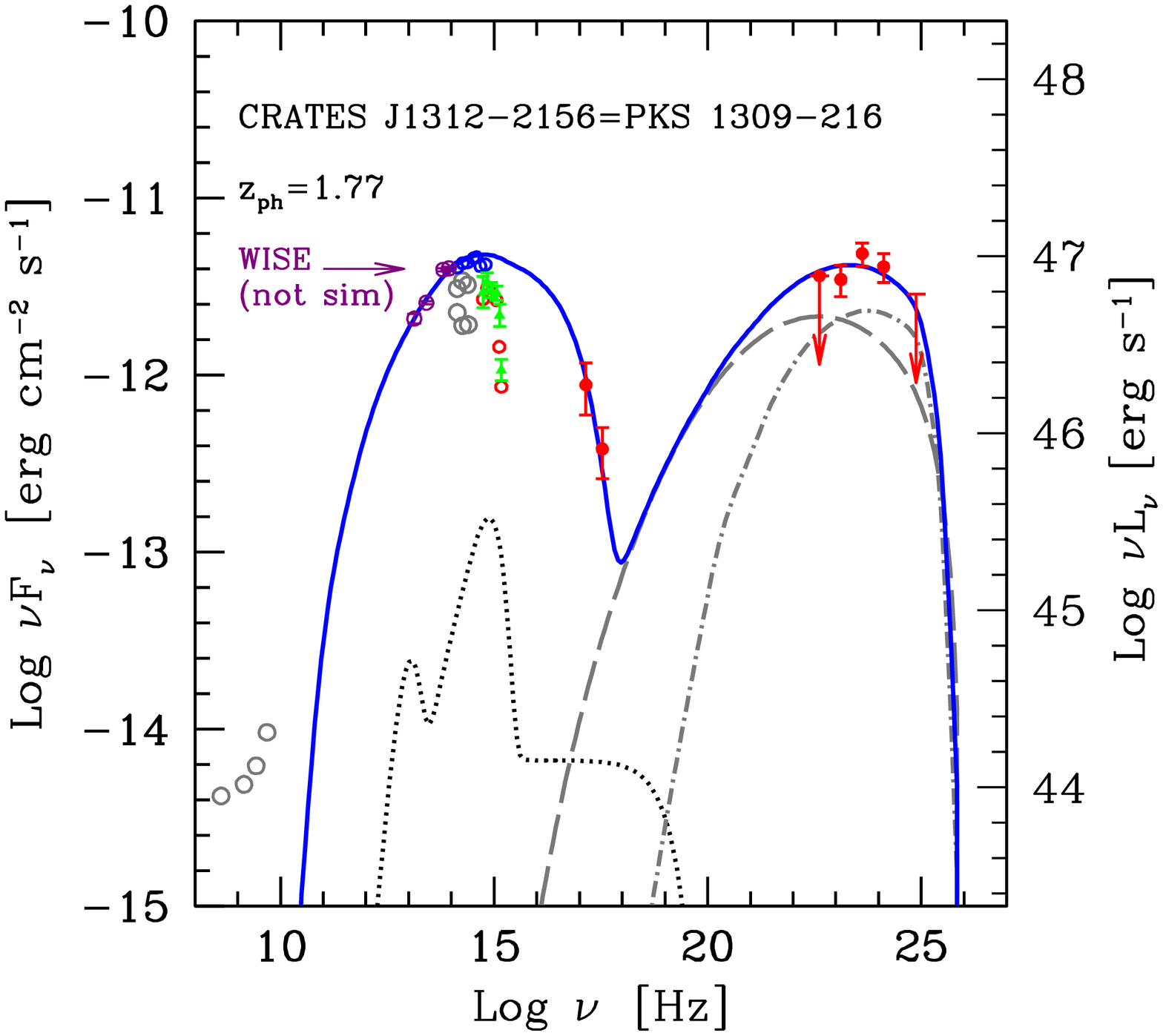,height=8cm,width=8.5cm}
\vskip -0.5cm
\caption{ 
Top panel: SED of RX J0537--5718 (top); CRATES 0630--2406 
(mid) and 1312--216 (bottom). 
The model assumes the presence of external radiation, as in Fig. \ref{0035}.
For CRATES 0630--2406 we also show a pure SSC model (thick black dashed line).
}
\label{1312}
\end{figure}

\section{Modeling}

The adopted model is a one--zone and leptonic model, fully
described in Ghisellini \& Tavecchio (2009).
The main properties are summarized in the Appendix, mainly to explain the 
meaning of the parameters listed in Tab. \ref{para} and Tab. \ref{powers}. 
In extreme synthesis, the model assumes that the bulk of the 
jet dissipation takes place
in one zone located at some distance $R_{\rm diss}$ from the black hole.
For simplicity, the emitting region is assumed spherical with a radius $R=\psi R_{\rm diss}$,
with $\psi=0.1$.  
The region is moving with a bulk Lorentz factor $\Gamma$, and is observed under a viewing
angle $\theta_{\rm v}$.
Energetic electrons are injected throughout the source for a time equal to the light crossing 
time $R/c$, and the particle distribution is calculated (though the continuity equation)
at this time, considering radiative losses
and possible electron--positron pair production and their reprocessing.

A specific feature of our modeling is that it takes into account
how the relevant energy densities change as a function of $R_{\rm diss}$,
the distance from the black hole of the main dissipation region.
Fig. \ref{u0035} illustrates this point, showing how the different contributions
to the energy density (as seen in the comoving frame) change as
a function of $R_{\rm diss}$ 
(see the lower axis), and, equivalently,
as a function of $R_{\rm diss}/R_{\rm S}$ (top axis), where
$R_{\rm S}$ is the \sch\ radius (for a black hole of mass $M=5\times 10^8 M_\odot$).
Since the jet is assumed to accelerate 
(with $\Gamma=\min[\Gamma_{\rm f}, (R_{\rm diss}/3R_{\rm S})^{1/2}]$),
the bulk Lorentz factor is relatively small in the inner regions of the jet.
As a consequence, the external radiation is not strongly boosted.
On the other hand, if the Poynting flux initially dominates the 
total power of the jet, the inner regions will be magnetically dominated.
As a result the main source of cooling is synchrotron and SSC.
When $\Gamma$ becomes large, but still $R_{\rm diss}<R_{\rm BLR}$,
the main source of cooling is the inverse Compton scattering off the
broad line photons (seen strongly enhanced in the comoving frame).
The adopted scaling between $R_{\rm BLR}$ and the disc luminosity implies that,
in the comoving frame, the external radiation energy density simply is:
\begin{equation}
U^\prime_{\rm BLR} \, \sim \, \Gamma^2 {L_{\rm BLR} 
\over 4\pi R_{\rm BLR}^2 c} \, =\, {\Gamma^2 \over 12\pi}\,\,\, {\rm erg\, cm^{-3}}
\label{ublr}
\end{equation}
where we have assumed that $L_{\rm BLR} =0.1 L_{\rm d}$ and  $R_{\rm BLR} =10^{17}L_{d,45}^{1/2}$ cm.
Beyond $R_{\rm BLR}$ this energy density decreases fast, because it is not enhanced
any longer by the Doppler boosting.
There is, however, another important external component, if the source has a molecular torus
re--emitting in the infrared the intercepted disc radiation.
Again, we assume a scaling between the size of the torus, $R_{\rm IR}$, and the
disc luminosity of the kind $R_{\rm IR} = 2.5\times 10^{18} L_{\rm d, 45}^{1/2}$ cm.
Assuming that the torus intercepts a fraction $a$ of the disc radiation, 
as long as $R_{\rm diss}<R_{\rm IR}$  the corresponding energy density 
measured in the comoving frame is:
\begin{equation}
U^\prime_{\rm IR} \, \sim \, \Gamma^2 {L_{\rm IR} \over 4\pi R_{\rm IR}^2 c} \, 
=\, 10^{-2}\, {a_{-1} \Gamma^2  \over 12\pi}\,\,\, {\rm erg\, cm^{-3}}
\label{uir}
\end{equation}
where $a_{-1}=a/10^{-1}$.
Comparing Eq. \ref{ublr} with Eq. \ref{uir} we see that, going from inside to
outside the BLR there is a ``jump"
of two orders of magnitudes in external radiation energy density,
as shown in Fig. \ref{u0035}, 
as $R_{\rm diss}$ moves from just inside to just outside $R_{\rm BLR}$.

Note that the magnetic energy density becomes comparable with
$U^\prime_{\rm BLR}$ and $U_{\rm IR}$ only in a narrow
range of distances, just beyond $R_{\rm BLR}$.

The different cooling rate reflects in a different electron energy
distribution $N(\gamma)$, having a break at a different $\gamma_{\rm c}$
defined as the random Lorentz factor of those electrons loosing half of
their energy in one dynamical time:
\begin{equation}
\gamma_{\rm c} = {3 m_{\rm e} c^2 \over 4\sigma_{\rm T} R U^\prime }
\label{gammac}
\end{equation}
If these are the electrons responsible for the peak of the SED,
we expect a large difference between blazars emitting inside or 
outside the BLR.
In fact, the jump in $U^\prime$ is only partly 
compensated by the increase of $R$.
Assume a change of a factor 3 in $R_{\rm diss}$ (between being inside and
being outside the BLR), and a corresponding change of a factor 3 in the 
magnetic field ($B\propto R^{-1}$). 
Since the jump in $U^\prime$ is nearly 100,
we expect that the peak synchrotron frequency $\nu_{\rm s}$ of
the region $R_{\rm out}$ is larger than the $R_{\rm in}$ one by a factor:
\begin{equation}
{\nu_{\rm s, out} \over \nu_{\rm s, in}}\sim 
{ B_{\rm out} \gamma^2_{\rm c, out} \over B_{\rm in} \gamma^2_{\rm c, in} } \, =\, 
{ B_{\rm out}\over B_{\rm in}}  
\left[ {R_{\rm in} U^\prime_{\rm in} \over R_{\rm out} U^\prime_{\rm out} } \right]^2
 \sim 300
\label{nunu}
\end{equation}
where the subscripts ``out" and ``in" stand for outside and inside the BLR.
It follows that if a ``standard" FSRQ with $R_{\rm diss}\lsim R_{\rm BLR}$ has
a synchrotron peak at $\nu_{\rm s}\sim 10^{13}$ Hz, then a blue FSRQ
-- with $R_{\rm diss}$ just a factor 3 larger 
(sufficient to make  $R_{\rm diss}>R_{\rm BLR})$ --
has a synchrotron peak at $\sim 3\times 10^{15}$ Hz, in the UV.
The example above is only illustrative, since the
different cooling will affect the shape of the 
particle distribution, not only its break,
but it does explain the expected difference
between ``red" and ``blue" FSRQs.

\begin{table*} 
\centering
\begin{tabular}{llllllllllllllll}
\hline
\hline
Name   &$z$ &$R_{\rm diss}$ &$M$ &$R_{\rm BLR}$ &$P^\prime_{\rm i}$ &$L_{\rm d}$ &$B$ &$\Gamma$ &$\theta_{\rm v}$
     &$\gamma_{\rm b}$ &$\gamma_{\rm max}$ &$s_1$  &$s_2$   &$\gamma_{\rm peak}$ &$U^\prime$\\ 
~[1]      &[2] &[3] &[4] &[5] &[6] &[7] &[8] &[9] &[10] &[11]  &[12] &[13]  &[14]   &[15]  &[16] \\
\hline   
0035 (EC)    &1.28 &300 (2000) &5e8 &173 &5.0e--4 &3.0 (0.04)  &1.28  &18 &1.5 &7e3   &2e5     &--1    &3     &7561   &0.1137 \\ 
0035 (SSC)   &1.28 &300 (2000) &5e8 &--- &7.0e--4 &---         &0.18  &24 &1.3 &1.5e4 &3e5     &1.6    &2.8   &1.18e4 &2.1e--3 \\ 
0035 (SSC)   &0.3  &180 (1200) &5e8 &--- &4.0e--4 &---         &0.08  &20 &2   &3.2e4 &5e5     &1.5    &4     &3.8e4  &4.4e--4 \\ 
0537--5718   &1.55 &390 (2600) &5e8 &229 &3.2e--3 &5.25 (0.07) &1.39  &12 &3   &6e3   &1.8e5   &--0.5  &2.7   &5583   &0.114   \\
0630--2406   &1.60 &375 (2500) &5e8 &194 &5.2e--3 &3.75 (0.05) &0.86  &17 &2.2 &1e4   &2.5e5   &0      &2.7   &1.16e4 &0.0553 \\ 
0630 (SSC)   &1.60 &312 (1300) &8e8 &--- &1.7e--3 &---         &0.38  &22 &1.5 &3e4   &2.5e5   &0.75   &3.    &2.07e4 &0.0102 \\ 
1312--216    &1.77 &480 (2000) &8e8 &245 &0.014   &6 (0.05)    &0.84  &14 &3   &3e3   &1e5     &--1    &2.2   &4484   &0.0601 \\ 
\hline
\hline 
\end{tabular}
\vskip 0.4 true cm
\caption{List of parameters used to construct the theoretical SED.
Not all of them are ``input parameters" for the model, because $R_{\rm BLR}$
is uniquely determined from $L_{\rm d}$.
Col. [1]: name;
Col. [2]: redshift;
Col. [3]: dissipation radius in units of $10^{15}$ cm and (in parenthesis) in units of \sch\ radii;
Col. [4]: black hole mass in solar masses;
Col. [5]: size of the BLR in units of $10^{15}$ cm;
Col. [6]: power injected in the blob calculated in the comoving frame, in units of $10^{45}$ erg s$^{-1}$; 
Col. [7]: accretion disc luminosity in units of $10^{45}$ erg s$^{-1}$ and (in parenthesis) in units of $L_{\rm Edd}$;
Col. [8]: magnetic field in Gauss;
Col. [9]: bulk Lorentz factor at $R_{\rm diss}$;
Col. [10]: viewing angle in degrees;
Col. [11] and [12]: break and maximum random Lorentz factors of the injected electrons;
Col. [13]: and [14]: slopes of the injected electron distribution [$Q(\gamma)$] below and above $\gamma_{\rm b}$;
Col. [15]: random Lorentz factors of the electrons radiating at the synchrotron peak;
Col. [16]: sum of the radiation and magnetic energy density in the comoving frame.
The total X--ray corona luminosity is assumed to be in the range 10--30 per cent of $L_{\rm d}$.
Its spectral shape is assumed to be always $\propto \nu^{-1} \exp(-h\nu/150~{\rm keV})$.
}
\label{para}
\end{table*}

\subsection{Physical parameters of blue FSRQs}

The results of the models are shown in Fig. \ref{0035} and Fig. \ref{1312}, and the 
model parameters are listed in Tab. \ref{para} and \ref{powers}.
We alert the reader that since our model assume a one--zone emitting region,
that must be quite compact, we cannot account for the radio emission,
that must be produced in much larger regions, not to be self--absorbed.
The observed self--absorbed synchrotron frequency of our model is typically
$\sim$1000 GHz.
Another concern is the value of the black hole mass.
The model could estimate it if the accretion disc
contribute to the optical--UV flux,
but in all cases we found instead that it is dominated by the 
non--thermal beamed continuum.
Therefore we simply assumed a black hole mass value
similar to the average value (i.e. $M\sim 5\times 10^8 M_\odot$)
found in Sbarrato et al. (2012).

\vskip 0.2 cm
\noindent
{\it RX J0035.2+1515 and CRATES J0630--2406 ---}
We have applied a SSC+EC model and a pure SSC model for both sources.
In the first case the model assumes the presence of a standard, Shakura--Sunyaev disc
and a reprocessing torus intercepting 10\% of the disc radiation, and reemitting
it in the IR.
The accretion disc luminosity has been chosen to agree with the limits given
in the previous section.
The radiation produced by the torus, together with the (not greatly boosted) BLR emission
is the dominant component for the inverse Compton scattering
producing the very high energy flux, while the SSC component gives a comparable
contribution in the X--ray band.
In these cases $R_{\rm diss}>R_{\rm BLR}$, but only by a factor 1.5--2.5.
The total (magnetic plus radiative) energy density as seen
in the comoving frame is small, and the energy of the electrons emitting
at the peaks of the SED is large.
The synchrotron peak frequency is beyond $10^{15}$ Hz for both sources,
and the synchrotron flux overwhelms the disc emission.
The magnetic field is not far from equipartition with the radiation energy density,
implying roughly equal synchrotron and Self Compton luminosities.

We have then applied a pure SSC model, to see if the presence of the
disc and torus is required or not. 
As can be seen, this model can account for the IR--$\gamma$--ray SED
quite satisfactorily, at the cost of increasing the Doppler factor $\delta$
(achieved by increasing $\Gamma$ and decreasing $\theta_{\rm v}$), and
lowering the magnetic field.

Finally, we have assumed $z=0.3$ for RX J0035.2+1515 and we have applied a pure SSC model,
to see if with this redshift the source could be fitted with the 
same parameters of high frequency BL Lacs.
Indeed, we find consistent values, very similar to other sources of this kind
(see Tavecchio et al. 2010; Ghisellini et al. 2010).
From the point of view of model fitting, we cannot
decide if RX J0035.2+1515 is a blue FSRQs or a high frequency peaked BL Lacs, but
in any case we require a large $\gamma_{\rm peak}$ and a small
comoving energy density (see the discussion below).

\vskip 0.2 cm
\noindent
{\it SUMMS J053748--571828 and CRATES 1312--2156 ---}
For these two sources we confirm the break in the bluest 
UVOT filters. 
In the absence of absorption, this break would be inconsistent with the extrapolation 
of the X--ray spectrum at lower frequencies.
We have applied only one model, assuming the presence of a standard accretion disc and torus,
and $z=z_{\rm ph}$.
With this redshift, the radio luminosity is large: this suggests that these sources
are powerful FSRQs.
Again, we have assumed a disc luminosity in agreement with the limits/indications
found in the previous section, a torus re--emitting 5--10\% of $L_{\rm d}$,
and $R_{\rm diss}$ nearly twice $R_{\rm BLR}$.
The bulk Lorentz factor ($\Gamma=12$) and viewing angle ($\theta_{\rm v}=3^\circ$) are
rather usual for {\it Fermi} FSRQs (Ghisellini et al. 2010).
The assumption of $R_{\rm diss}>R_{\rm BLR}$ implies a large $\gamma_{\rm peak}$ and a small
comoving energy density.

\begin{figure}
\vskip -0.8cm
\hskip -0.3cm
\psfig{file=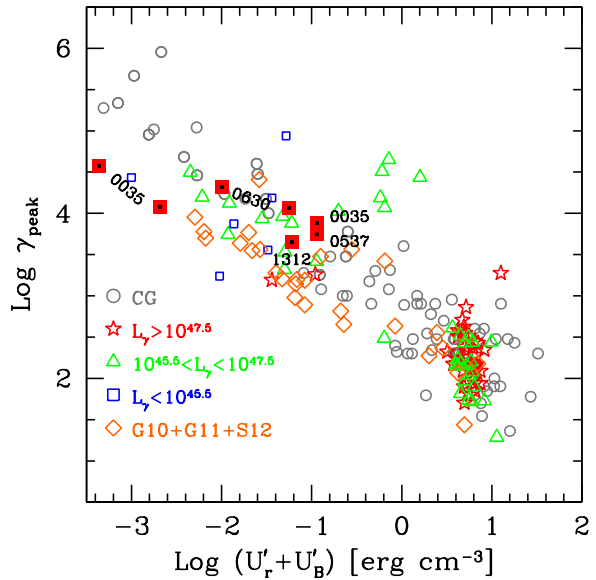,height=9cm,width=9cm}
\vskip -0.5 cm
\caption{
The Lorentz factor of the electrons emitting at the synchrotron peak, $\gamma_{\rm peak}$,
as a function of the radiation plus magnetic energy density, as seen in the comoving frame.
The 5 sources discussed in this paper are labelled and are compared to a collection
of blazar samples, discussed in Celotti \& Ghisellini (2008) (CG); in 
Ghisellini et al. (2010, 2011) and in Sbarrato et al. (2012) (G10+G11+S12).
As can be seen, the ``blue quasars" studied here lie along the main trend
of decreasing $\gamma_{\rm peak}$ when increasing the energy density.
}
\label{gammap}
\end{figure}

\vskip 0.2 cm
Concerning the jet power (see Tab. \ref{powers}), 
our sources have values similar to the other {\it Fermi} blazars
(studied in Ghisellini et al. 2010), but certainly not extreme, at the border between
BL Lacs and FSRQs.
Consider that since the average electron energy is large, there is little difference
in considering the presence of one (cold) proton per emitting electron or not.
Even a pure electron--positron pair jet would carry a similar total jet power.
 
The main finding of the modeling is that, in order to obtain a good agreement
with the data, a large $\gamma_{\rm peak}$ and a small energy density $U^\prime_{\rm B} +U^\prime_{\rm r}$
is required {\it in any case,} 
either if the model is a pure SSC or if it includes
external radiation.
Fig. \ref{gammap} compares the obtained values of 
$\gamma_{\rm peak}$ and $U^\prime_{\rm B} +U^\prime_{\rm r}$
with the ones of other $\gamma$--ray detected blazars.
As can be seen, all our sources follow the same trend of other blazars.
This implies that they are consistent with the explanation of the blazar sequence based 
on radiative cooling, as suggested in Ghisellini et al. (1998).

Rau et al. (2012) considered about 100 blazars, finding an upper limit on the redshift
for the majority of them, and a photometric value for 11.
Of the latter, only 4 (the ones considered by P12) have a flat optical shape 
(spectral index $\alpha<1$, $F(\nu)\propto \nu^{-\alpha}$).
The remaining blazars are then either closer or steeper.
Therefore blue quasars (powerful blazars with $\alpha<1$) are rare objects,
as predicted in Ghisellini \& Tavecchio (2008) on the basis of the small parameter space
occupied by these sources. 
Consider also that if RX J0035.2+1515 is indeed at more
moderate redshifts, it could become a high frequency peaked BL Lac,
implying that blue quasars are even rarer.
In turn this implies, in the framework of our model, that the dissipation region
in the jet is almost always within the BLR.

\begin{table} 
\centering
\begin{tabular}{lllll}
\hline
\hline
Name   &$\log P_{\rm r}$ &$\log P_{\rm B}$ &$\log P_{\rm e}$ &$\log P_{\rm p}$ \\
\hline   
0035 ($z$=1.28, EC)  &44.20  &45.25  &43.09  &43.66  \\  
0035 ($z$=1.28, SSC) &44.05  &43.80  &44.26  &45.72 \\
0035 ($z$=0.3, SSC)  &42.91  &42.49  &43.61  &44.59  \\
0537--5718           &44.65  &45.20  &43.48  &44.12  \\  
0630--2406           &45.16  &45.05  &44.12  &44.56  \\  
0630 (SSC)           &44.85  &44.41  &44.28  &44.39  \\
1312--216            &45.42  &45.08  &44.42  &45.02  \\  
\hline
\hline 
\end{tabular}
\vskip 0.4 true cm
\caption{
Logarithm of the jet power in the form of radiation ($P_{\rm r}$), 
Poynting flux ($P_{\rm B}$),
bulk motion of electrons ($P_{\rm e}$) and protons ($P_{\rm p}$,
assuming one proton per emitting electron). 
Powers are in erg s$^{-1}$.
}
\label{powers}
\end{table}

\begin{figure*}
\vskip -1.3cm
\hskip 0.2 cm
\psfig{file=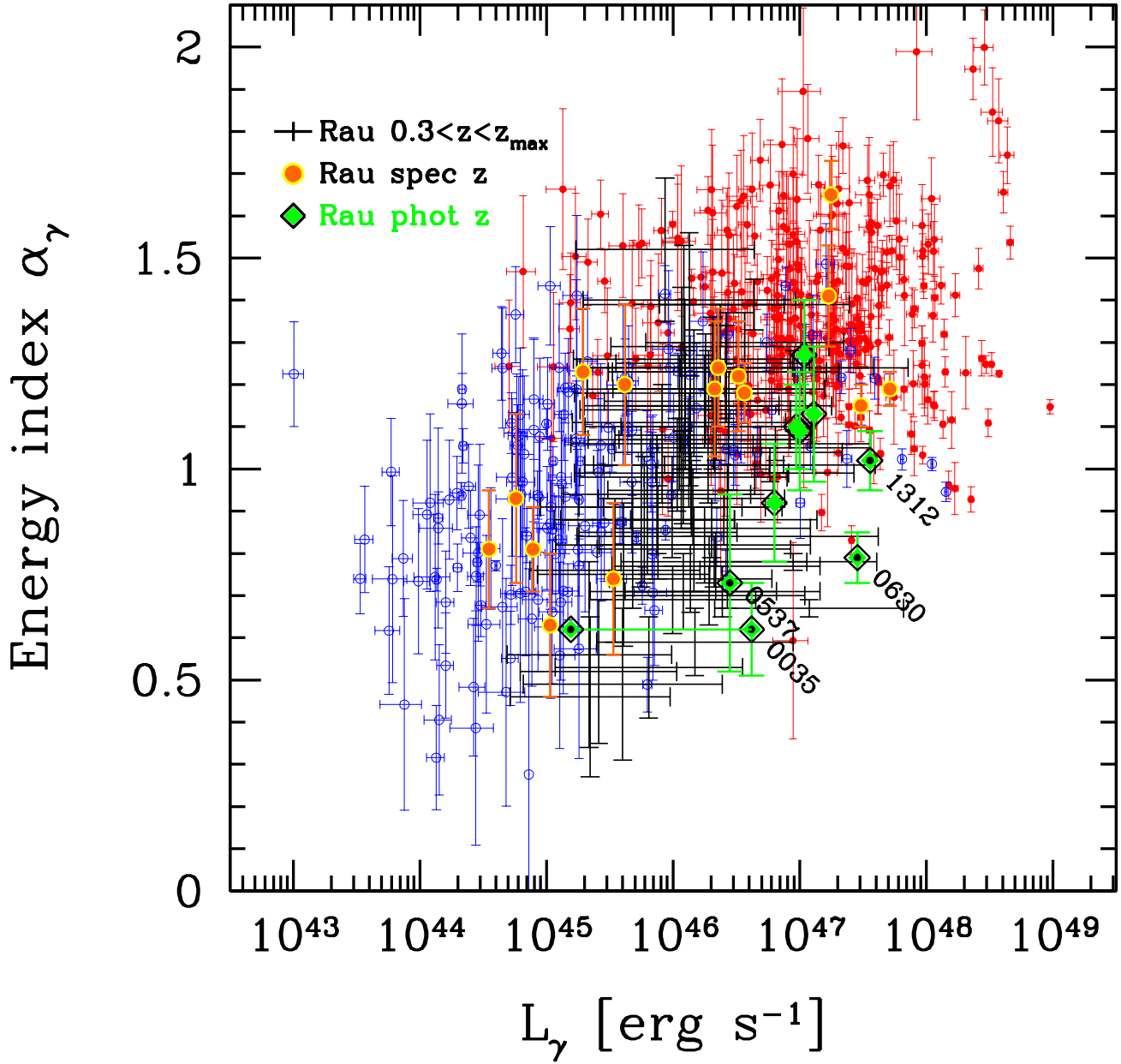,height=18.8cm,width=18.8cm}
\vskip -0.7 cm
\caption{
The energy spectral index $\alpha_\gamma$ as measured by {\it Fermi}
as a function of their [0.1--10 GeV] (rest frame) luminosity.
Filled (red in the electronic version) points are FSRQs, empty (blue) circles
are BL Lacs.
The sources studied in R12 are plotted with larger symbols
if the redshift is determined (spectroscopically or photometrically).
Sources with only un upper limit for the photometric redshifts
are plotted assuming a luminosity the upper limit as the maximum redshift,
and $z_{\rm min}=0.3$ as the minimum (this corresponds approximately
to the limit at which the host galaxy should be visible). For 
RX J0035.2+1515 we  plot also the location
of the source assuming $z=0.3$.
}
\label{divide}
\end{figure*}

\section{The updated Fermi blazars' sequence}

Ghisellini, Maraschi \& Tavecchio (2009) pointed out
the existence of a well defined trend in the $\alpha_\gamma$--$L_\gamma$
plane of the blazars detected after the first three months of {\it Fermi} 
observations. 
BL Lacs and FSRQs were separated in this plane, with BL Lacs having the
flattest slopes. 
These were very bright blazars, detected at more than the $10\sigma$ level.
Now the 2 years catalog is available, and we can compare the locations
of the R12 sources in the updated version of the $\alpha_\gamma$--$L_\gamma$
plane. 
Fig. \ref{divide} shows the results considering all BL Lacs
(blue empty points) and all FSRQs (filled red circles) with redshift
present in the clean 2LAC sample.
What plotted is the [0.1--10 GeV] K--corrected luminosity.
Overplotted to these sources, we show the blazars considered by R12 
that have spectroscopic and photometric redshifts, together with
the blazars for which the photometric analysis yields only 
a {\it upper limit} on the redshift ($z<z_{\rm max}$).
For the latter sources, we have considered a range of redshifts,
between 0.3 and $z_{\rm max}$, hence a range of $L_\gamma$.
The choice of $0.3$ as the minimum possible redshift follows from the
fact that, at smaller redshifts, the host galaxy starts to be visible.
Note the following:

\begin{itemize}

\item 
Comparing this figure with the analogous figure in 
Ghisellini, Maraschi \& Tavecchio (2009),
the trend between $\alpha_\gamma$ and $L_\gamma$ 
is now even more clear.
Likely, this is due to the improved software analysis, 
assigning to the background more low energy photons than before.
As a result, steep spectra ($\alpha_\gamma>1$) become flatter, while already
flat sources are less affected.
We recall that the observed $\alpha_\gamma$--$L_\gamma$ trend is partly due to selection effects:
when deeper surveys will be available, we will start to
see slightly misaligned blazars of lower $L_\gamma$ (and same $\alpha_\gamma$)
as well as blazars with smaller black hole masses (and presumably
with smaller $L_\gamma$), so that the trend will be much more dispersed.

\item 
R12 consider some blazars with spectroscopic redshift,
but with no redshift indication in the 2LAC catalog.
These are plotted with different symbols in Fig. \ref{divide}.
There are two groups: five blazars have a flat $\alpha_\gamma$
and are candidates to be high frequency peaked BL Lacs.
The remaining ones are located in the region populated by FSRQs,
and are then candidates to be normal FSRQs.

\item
The blazars with the estimated photometric redshift have all large $L_\gamma$.
The 4 blazars studied here are the flattest.
Remember that they were selected to have a flat optical spectrum:
they also have a flat $\gamma$--ray spectrum.
Fig. \ref{divide} shows how the blazar RX J0035.2+1515
would be located if its redshift were $z=0.3$.
In this case it would be a normal highly peaked BL Lac object.

\item
The blazars in R12 with only a limit on $z$ are located in the 
region of the plot already populated by sources with known $z$.
There are no outliers.
The existence of very powerful BL Lac objects
with flat $\gamma$--ray spectra is excluded.
This agrees with the blazar sequence.

\end{itemize}

We can conclude that the observed boundary in the $\alpha_\gamma$--$L_\gamma$ 
plane at large luminosities is not due to selection effects, namely to
the difficulty to determine the redshifts of powerful BL Lac objects.
It must have a physical origin.
We (Ghisellini, Maraschi \& Tavecchio 2009) have proposed that this is due to 
blazars with the largest black hole masses, accreting at different rates. 

Small accretion rates correspond to weak discs:
the ionizing luminosity $L_{\rm ion}$ is small, the broad line region is correspondingly
weak and very close to the black hole (if $R_{\rm BLR}\propto L_{\rm ion}^{1/2}$
still holds).
Dissipation takes place beyond the BLR, the radiative cooling is weak, 
electrons reach high energies, and from an observational point of view we 
have a high frequency peaked BL Lac with no or very weak broad lines 
(being the synchrotron peak in the UV/X-ray, it is easier for the jet emission to 
overwhelm the weak disc and BLR emission).
Increasing the accretion rate, we enter in the regime of radiatively efficient discs
and then a normal BLR.
If the jet dissipates mainly within the BLR, the radiative cooling is strong,
implying a low frequency peaked FSRQs.
Furthermore, if the jet power is proportional to the accretion rate $\dot M$,
then $L_\gamma$ is also linked with it, and therefore we have the 
chain: low $\dot M \to$ low jet power + weak and small BLR $\to$ flat $\alpha_\gamma$
+ small $L_\gamma$.
And the opposite for large $\dot M$.
This chain can be repeated for each black hole mass.
Of course, the lowest the black hole mass, the lowest the maximum $L_\gamma$.

\section{Conclusions}

We have studied the nature of a few blazars, whose redshift
has been recently determined, photometrically, by Rau et al. (2012).
In addition, we have considered all blazars for which
Rau et al. (2012) could establish a lower limit on the redshift.
Our conclusions agree with Padovani et al. (2012): the 4 blazars
considered in our and their papers are probably FSRQs (with a residual
doubt about the determination of the photometric redshift
of RX J0035.2+1515, because the UVOT data analysis is complex).

Independently on the actual redshift of RX J0035.2+1515, all 4 blazars
are ``blue" (i.e. the synchrotron and the inverse Compton peak frequencies are large)
because the radiative cooling is weak.
If they do have broad lines, they are overwhelmed by
the strong synchrotron emission.
This agrees completely with the standard explanation of the blazar sequence.
The fact that it was possible to estimate a photometric redshift for a few sources,
out of a sample of nearly one hundred, confirms that these blue FSRQs are rare.
This in turn implies that it is rare, for a jet, to produce most of the radiation
we see at large distances from its black hole.

The found blue quasars  can
have accretion discs emitting at a level of a few per cent of the
Eddington one, ``normal" black hole masses (namely between $10^8$ and $10^9 M_\odot$)  
and bulk Lorentz factors.
Their properties are similar to the average properties of other blazars.

We have considered the updated $\alpha_\gamma$--$L_\gamma$ plane, considering all sources  
with spectroscopic redshifts within the 2LAC sample and including the 
blazars for which Rau et al. (2012) could estimate an upper limit on $z$.
We find no outliers in this plane: in other words, the upper limits on the
redshift exclude the presence of very powerful BL Lacs.
This confirms our previous explanation of the separation of BL Lacs and FSRQs in this plane
based on a sequence of accretion rates.


\section*{Acknowledgements}
This research has made use of the NASA/IPAC Extragalactic Database (NED) which is 
operated by the Jet Propulsion Laboratory, California Institute of Technology, under 
contract with the National Aeronautics and Space Administration. 
Part of this work is based on archival data software or on--line services provided by the ASI
Data Center (ASDS).
This publication makes use of data products from the Wide-field Infrared Survey Explorer, 
which is a joint project of the University of California, Los Angeles, and the 
Jet Propulsion Laboratory/California Institute of Technology, funded by the 
National Aeronautics and Space Administration.


\vskip 0.5 cm
\noindent
{\bf APPENDIX}
\vskip 0.3 cm
\noindent
At a distance $R_{\rm diss}$ from the black hole of mass $M$
the jet dissipates
part of its power and injects relativistic electrons throughout the
emitting region, assumed to be spherical, with radius $R=\psi R_{\rm diss}$,
with $\psi=0.1$.  
In the region there is a tangled magnetic field $B$.
The relativistic electrons are injected with a smoothly joining broken power law
in energy: 
\begin{equation}
Q(\gamma)  \, = \, Q_0\, { (\gamma/\gamma_{\rm b})^{-s_1} \over 1+
(\gamma/\gamma_{\rm b})^{-s_1+s_2} } \quad {\rm [cm^{-3} s^{-1}]} 
\label{qgamma}
\end{equation}
The energy particle distribution $N(\gamma)$ [cm$^{-3}$] is calculated
solving the continuity equation where particle injection, radiative
cooling and pair production (via the $\gamma$--$\gamma \to e^\pm$
process), are taken into account. 
The created pairs contribute to the emission.

The injection process lasts for a light crossing time $R/c$, and we
calculate $N(\gamma)$ at this time.  This assumption comes from the
fact that even if injection lasted longer, adiabatic losses caused by
the expansion of the source (which is traveling while emitting) and
the corresponding decrease of the magnetic field would make the
observed flux to decrease.  Therefore the calculated spectra
correspond to the maximum of a flaring episode.

The total power injected into the source in the form of relativistic
electrons is $P^\prime_{\rm i}=m_{\rm e}c^2 V\int Q(\gamma)\gamma
d\gamma$, where $V=(4\pi/3)R^3$ is the volume of the emitting region.

The bolometric luminosity of the accretion disc
is $L_{\rm d}$.
Above and below the accretion disc, in its inner parts, there is an
X--ray emitting corona of luminosity $L_{\rm X}$ (it is fixed at a
level of 30\% of $L_{\rm d}$).  Its spectrum is a power law of energy
index $\alpha_X=1$ ending with a exponential cut at $E_{\rm c}=$150
keV.  The specific energy density (i.e. as a function of frequency) of
the disc and the corona are calculated in the comoving frame of the
emitting blob, and used to properly calculate the resulting External
inverse Compton spectrum.  The BLR is assumed to be a thin spherical
shell, of radius $R_{\rm BLR}=10^{17} L_{\rm d, 45}^{1/2}$ cm.
We consider also the presence of a IR torus, at larger distances.  The
internally produced synchrotron emission is used to calculate the
synchrotron self Compton (SSC) flux.  Table \ref{para} lists the
adopted parameters.

The power carried by the jet can be in the form of
radiation ($P_{\rm r}$), magnetic field ($P_{\rm B}$), emitting
electrons ($P_{\rm e}$, no cold electron component is assumed) and
cold protons ($P_{\rm p}$, assuming one proton per emitting electron).
All the powers are calculated as
\begin{equation}
P_i  \, =\, \pi R^2 \Gamma^2\beta c \, U^\prime_i
\end{equation}
where $U^\prime_i$ is the energy density of the $i$ component, as
measured in the comoving frame.

The power carried in the form of the produced radiation, $P_{\rm r}
=\pi R^2 \Gamma^2\beta c \, U^\prime_{\rm rad}$, can be re--written as
[using $U^\prime_{\rm rad}=L^\prime/(4\pi R^2 c)$]:
\begin{equation}
P_{\rm r}  \, =\,  L^\prime {\Gamma^2 \over 4} \, =\, L {\Gamma^2 \over 4 \delta^4}
\, \sim \, L {1 \over 4 \delta^2}
\end{equation} 
where $L$ is the total observed non--thermal luminosity ($L^\prime$ is
in the comoving frame) and $U^\prime_{\rm rad}$ is the radiation
energy density produced by the jet (i.e.  excluding the external
components).  The last equality assumes $\theta_{\rm v}\sim 1/\Gamma$.

When calculating $P_{\rm e}$ (the jet power in bulk motion of emitting
electrons) we include their average energy, i.e.  $U^\prime_{\rm e}=
n_{\rm e} \langle\gamma\rangle m_{\rm e} c^2$.

\end{document}